\documentclass[acmlarge, nonacm]{acmart} 
\pdfoutput=1 
\usepackage[utf8]{inputenc}
\usepackage{enumitem}
\usepackage[]{mdframed}
\usepackage{caption}
\usepackage{subcaption}
\usepackage[T1]{fontenc}

\newcommand{\squishlist}{\begin{itemize}[itemsep=1pt,parsep=2pt,topsep=3pt,partopsep=0pt,leftmargin=0em, itemindent=1em,labelwidth=1em,labelsep=0.5em]}
\newcommand{\squishend}{\end{itemize}}

\AtBeginDocument{%
  \providecommand\BibTeX{{%
    \normalfont B\kern-0.5em{\scshape i\kern-0.25em b}\kern-0.8em\TeX}}}

\setcopyright{acmcopyright}
\copyrightyear{2023}
\acmYear{2023}
\acmDOI{XXXXXXX.XXXXXXX}

\acmJournal{IMWUT}
\acmVolume{1}
\acmNumber{1}
\acmArticle{1}
\acmMonth{1}

\begin{document}

\title[LLMs for Embedded System Development and Debugging]{Exploring and Characterizing Large Language Models for Embedded System Development and Debugging}

\author{Zachary Englhardt}
\email{zacharye@cs.washington.edu}
\orcid{0000-0002-6646-6466}
\affiliation{
  \institution{University of Washington}
  \country{USA}
}
\author{Richard Li}
\affiliation{
  \institution{University of Washington}
  \country{USA}
}
\author{Dilini Nissanka}
\affiliation{
  \institution{University of Washington}
  \country{USA}
}
\author{Zhihan Zhang}
\affiliation{
  \institution{University of Washington}
  \country{USA}
}
\author{Girish Narayanswamy}
\affiliation{
  \institution{University of Washington}
  \country{USA}
}
\author{Joseph Breda}
\affiliation{
  \institution{University of Washington}
  \country{USA}
}
\author{Xin Liu}
\affiliation{
  \institution{University of Washington}
  \country{USA}
}

\author{Shwetak Patel}
\affiliation{
  \institution{University of Washington}
  \country{USA}
}

\author{Vikram Iyer}
\email{vsiyer@uw.edu}
\orcid{0000-0002-3025-7953}
\affiliation{
  \institution{University of Washington}
  \country{USA}
}

\renewcommand{\shortauthors}{Englhardt et al.}


\begin{abstract}
Large language models (LLMs) have shown remarkable abilities to generate code, however their ability to develop software for embedded systems, which requires cross-domain knowledge of hardware and software has not been studied. In this paper we develop an extensible, open source hardware-in-the-loop framework to systematically evaluate leading LLMs (GPT-3.5, GPT-4, PaLM 2) to assess their capabilities and limitations for embedded system development. We observe through our study that even when these tools fail to produce working code, they consistently generate helpful reasoning about embedded design tasks. We leverage this finding to  study how human programmers interact with these tools, and develop an human-AI based software engineering workflow for building embedded systems. 

Our evaluation platform for verifying LLM generated programs uses sensor actuator pairs for physical evaluation. We compare all three models with N=450 experiments and find surprisingly that GPT-4 especially shows an exceptional level of cross-domain understanding and reasoning, in some cases generating fully correct programs from a single prompt. In N=50 trials, GPT-4 produces functional I2C interfaces 66\% of the time. GPT-4 also produces register-level drivers, code for LoRa communication, and context-specific power optimizations for an nRF52 program resulting in over 740x current reduction to 12.2\textmu A. We also characterize the models' limitations to develop a generalizable human-AI workflow for using LLMs in embedded system development. We evaluate our workflow with 15 users including novice and expert programmers. We find that our workflow improves productivity for all users and increases the success rate for building a LoRa environmental sensor from 25\% to 100\%, including for users with zero hardware or C/C++ experience.

\end{abstract}

\maketitle
\section{Introduction}
Large language models (LLMs) such as GPT-3.5, GPT-4 and PaLM 2 have recently made significant strides in code generation for a variety of software development tasks. These LLMs, which have been trained on many code samples, can generate syntactically correct and semantically meaningful code from high-level task descriptions, and rival human experts on certain tasks~\cite{bubeck2023sparks}. Researchers have also begun exploring optimizations such as encoding user intent~\cite{lahiri2022interactive}, self-repair for LLMs to correct their mistakes~\cite{olausson2023demystifying}, and collaboration between models~\cite{dong2023selfcollaboration}. Commercial integrated development environments (IDEs) have already begun to incorporate these tools~\cite{copilot1} suggesting AI assistants will play a growing role in software development workflows in years to come.

\begin{figure}[!t]
\centerline{\includegraphics[width=0.9\textwidth]{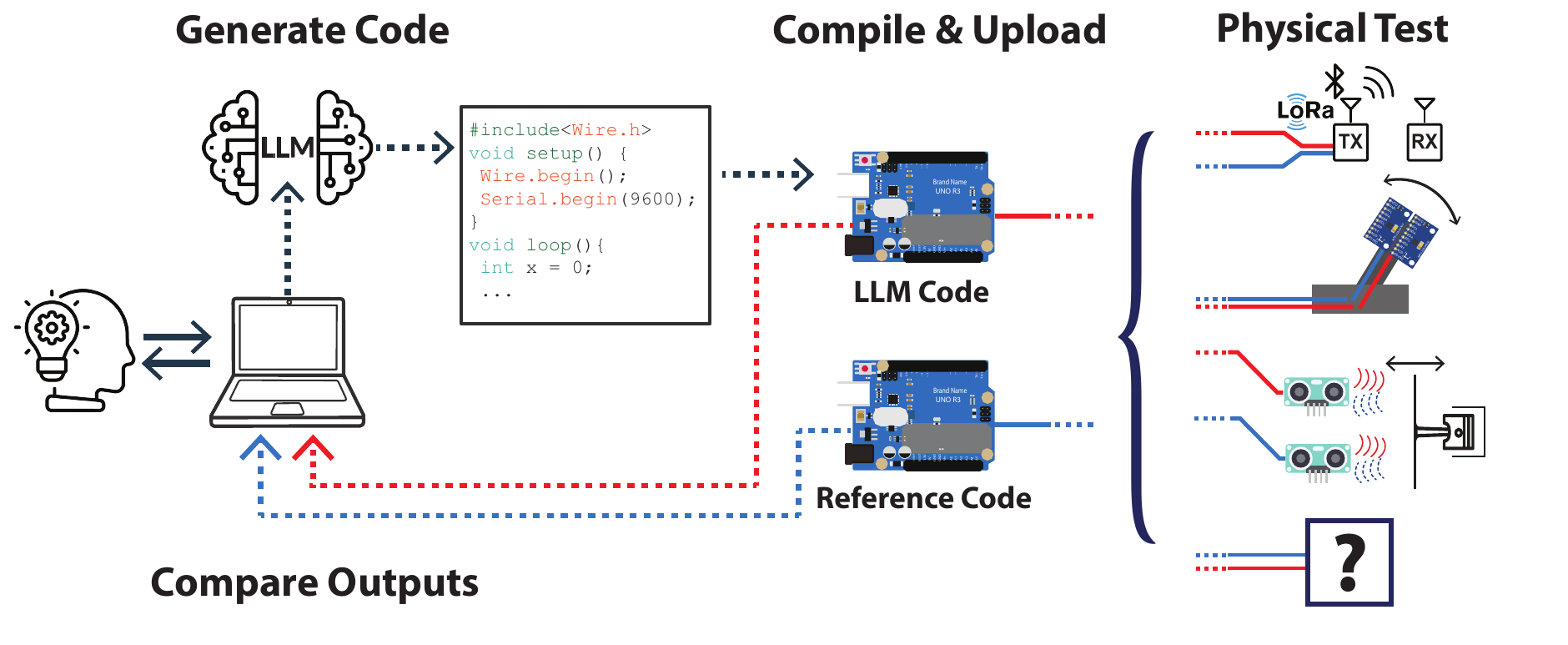}}
\vskip -0.2in
\caption{Illustration of human AI co-design for embedded system development with our automated hardware-in-the-loop testing framework. Our end-to-end pipeline uploads LLM generated code to a microcontroller, physically tests sensor actuator pairs against reference code, and evaluates the output. Our open framework can be extended to incorporate new sensors.}
\label{fig:overview}
\vskip -0.2in
\end{figure}

Despite these efforts, there has been little to no study of using LLMs for embedded systems development. While tools like Github Co-Pilot can auto-complete code, they have no link to the hardware, no context about the interconnections of components in the system, or what outputs they would produce. Embedded systems design and programming straddles the hardware-software interface and requires a cross-domain understanding of how devices \textit{interact with the physical world}~\cite{wolf2010high,lee2011introduction}. 
Development of embedded systems requires a cyclical process of continuous development and verification, entailing numerous software-hardware integration trials. This iterative approach differs from conventional programming, where outputs are deterministic and predictable. Embedded system development demands complex, multi-step reasoning and verification to ensure system reliability and functionality. For example, designing a simple wireless temperature sensor requires a physical understanding of the chips and their interfaces, network connectivity, and the sensed signals. These challenges impose a high barrier even for human programmers and have inspired projects such as the MakeCode platform focused specifically on improving accessibility to embedded systems development~\cite{karvinen2019lowering,devine2019makecode}. This raises a key question: can LLMs generate reasoning about these concepts, and how can they play an effective role in embedded system development?

In this work we conduct an exploratory analysis of leading large-language models (GPT-3.5 \cite{gpt-3.5}, GPT-4 \cite{openai2023gpt4}, and PaLM 2~\cite{anil2023palm}) on embedded system development and debugging tasks. We design physical interfaces for LLMs to directly interact with physical hardware and create an end-to-end hardware-in-the-loop (HIL) platform for evaluating LLM generated code. We aim to understand their capabilities and develop actionable insights toward human-AI co-development workflows with hardware-in-the-loop AI tools as shown in Figure~\ref{fig:overview}. 

Using LLMs for embedded systems development raises fundamental questions and challenges. First, how can we repeatably verify LLM generated code? Repeated trials are essential to advance our understanding in this domain beyond blog posts with a handful of experiments; however unlike programs with text output that can be verified, we seek to probe LLMs' abilities to generate code for devices like sensors that interact with the physical world. This requires a inducing and measuring real-world stimuli in the presence of environmental noise. A robust benchmarking pipeline is also the first step toward model optimization through fine-tuning and self-repair techniques with hardware in the loop.

Second, using LLMs with embedded systems raises fundamental questions about the ability of these models to reason about the interaction between the physical and digital environments where systems are deployed. The simple task of "Write a program that blinks an LED" requires an understanding of how components are connected (the correct output pin) and that setting the pin value to a logic level high will produce light. Given task-specific code-generation models, such as Github Copilot are not developed for the domain of embedded systems, it is unclear whether they can effectively engage with the intricacies of embedded systems interfacing with the physical world. General-purpose models may have a robust representation of these topics or the cross-domain reasoning ability to solve problems in this space.



To address these challenges we begin by developing methods to physically interface microcontroller and sensor outputs with LLMs to create an HIL evaluation platform for systematically verifying programs using sensor actuator pairs. We compare GPT-3.5, PaLM2 and GPT-4 with N=450 experiments and find surprisingly that GPT-4 especially shows  an exceptional level of cross-domainreasoning, in some cases generating fully correct programs from a single prompt. We perform a series of systematic evaluations probing a variety of embedded system development capabilities and find it generates register-level drivers, I2C interfaces, LoRa communication code, and even suggests specific and actionable methods to power optimize code. We make the surprising discovery that the models can also suggest context-specific hardware debugging advice like checking wiring.

We also observe and systematically document a number of limitations and errors. While the best model only produced perfect end-to-end code on the most complex tasks 14\% of the time (GPT-4, N=50 trials) these partially correct programs contained significant functional code, as well as detailed comments and explanation of how to correctly design the system. These findings suggest that while LLMs may play a role in fully automated code generation, perhaps more powerful is a new human-AI co-development approach to embedded systems development that combines human context and high level design objectives with LLM code generation. We synthesize these findings into a generalizable workflow to enable human-AI embedded system co-development. We evaluate this workflow through a user study with 15~participants and find the most dramatic benefits are to novice users. Bridging knowledge gaps of low level hardware implementation and personalized debugging advice allowed first time embedded programmers to design fully functional systems in 40~min. This human-AI approach has the potential to transform embedded systems education and help reduce the barrier for researchers across the Ubicomp community to work on hardware and physical computing topics.

We summarize our contributions below: 
\squishlist
    \item \textbf{Physical Testbench.} We build the first open-source, extensible software-hardware evaluation framework for Human-AI embedded system development. It includes an end-to-end pipeline for real-time LLM-powered embedded code validation with sensor-actutor pairs. We support state-of-the-art LLMs in our framework, including PaLM-2, GPT-3.5, and GPT-4. We use this setup to execut a comprehensive suite of 450 real-world evaluations involving integrated software-hardware interactions. We envision our framework will serve as a valuable asset for the research community, fostering further exploration and validation of sensor-driven software-hardware systems in ubiquitous computing.

    \item \textbf{Exploration of LLM capabilities and limitations.} We perform the first systematic evaluation of LLM facilitated embedded tasks and show they have a rich cross-domain understanding of hardware and software. We evaluate multiple microcontroller platforms including the Atmel ATMega328P (Arduino) and the Nordic nRF52832 (using the Nordic SDK in standard C). We find that LLMs can even provide specific and actionable hardware debugging advice about wiring and analyze programs, reducing power consumption on an nRF52 by over 740x to 12.2 \textmu A. We also characterize limitations like hallucinations and representing the system's state.
    \item \textbf{LLM-Based Human-AI development workflow.} We observe that LLMs produce detailed comments and explanations of how to design embedded systems as well as context-specific debugging advice. Leveraging these findings, we develop a set of prompting strategies and propose an AI integrated development workflow for embedded systems design, programming and debugging.
    \item \textbf{User Study.} We evaluate our workflow with 15 users across experience levels. We find our workflow improves productivity for users across all experience levels and increases the success rate on complex tasks from 25\% to 100\%. This allowed users with zero hardware or C/C++ experience to build a fully functional LoRa environmental sensor transmitter and receiver in 40~min, demonstrating the potential of this approach for education and lowering the barrier for working with hardware.

\squishend


\section{Related Work}
\textbf{Capabilities of Large Language Models.} State-of-the-art LLMs~\cite{openai2023gpt4, chowdhery2022palm, touvron2023llama} make use of transformer-based neural architectures consisting of billions to trillions of parameters \cite{vaswani2017attention}. These models, trained on internet-scale text corpora, have demonstrated the ability to perform logic-based reasoning, solve complex problems, have scored highly on a number of language-task benchmarks, and produce text outputs indistinguishable from that of human authors~\cite{bubeck2023sparks, kojima2205large, wang2019superglue}. These capabilities span several specialized domains wherein LLMs have demonstrated their efficacy as dependable foundational models. These domains encompass consumer health, medicine, education, finance, chip design, and environmental science, among others ~\cite{brown2020language, liu2023large, kojima2205large, liu2023chipnemo, wu2023bloomberggpt, rillig2023risks, singhal2023towards}. Such versatility highlights the potential of LLMs to serve as integral components within interdisciplinary research and application development.

\textbf{Tool Usage.} The capabilities of these models can be further extended by allowing LLMs access to APIs or tools, and providing these models with an interface to these peripheries. For example, prior work has shown that although LLMs are currently limited by their ability to do arithmetic, this may be alleviated through access to a calculator API~\cite{bubeck2023sparks}. Recent work has shown the utility of fine-tuning to improve interfacing with tooling APIs.~\cite{patil2023gorilla}. This highlights potential for developing physical tool interfaces for LLMs.

\textbf{Prompting Language Models.} While these LLMs may be powerful tools, research has found that high-quality prompting is required to receive consistently high-quality model outputs~\cite{bubeck2023sparks}. A number of works have developed frameworks and workflows to optimize LLM responses~\cite{white2023prompt, wei2022chain, wang2022self, yao2023tree, yao2022react, zhou2022large}. In large part these prompting techniques include using highly detailed, unambiguous prompts, providing additional context, and asking the model to self-evaluate.

\textbf{LLMs for Code Generation and Embedded Systems.} As LLMs are predominantly trained on internet-scraped text corpora, comprised, in-part, of open-source software projects (as found on Github and similar sites), these models encode a rich understanding of programming languages, code, and software flows. Prior work has shown the use of language models and AI assistants (e.g., Github Copilot) to generate software, with recent language models performing with the aptitude of competent software developers~\cite{chen2021evaluating, jain2022jigsaw, bubeck2023sparks, devlin2017robustfill, xu2022systematic, ziegler2022productivity, liu2023your, nguyen2022empirical, gunasekar2023textbooks}. A number of works have additionally fine-tuned LLMs specifically for code-generation across a number of programming languages showing improvement over the base models~\cite{nijkamp2022codegen, nijkamp2023codegen2, chen2023improving}. Researchers have also developed a number of methodologies to better prompt LLMs to generate code including self-debugging~\cite{chen2023teaching, shinn2023reflexion}, task decomposition~\cite{yao2022react, kim2023language}, and prompting frameworks~\cite{poesia2022synchromesh}. While prior work has evaluated LLM-based code generation on a number of programming languages, including C/C++ (commonly used for embedded software), these works do not explicitly evaluate the use of LLMs for embedded systems. Existing work on hardware-platform code-generation primarily revolves around tools such as IDEs, programming frameworks and devices to make hardware-level programming more approachable for novices but have not used LLMs with hardware~\cite{ball2019microsoft, devine2018makecode, koopman2005undergraduate, pasricha2022embedded, brennan2022exploring}. Other systems have focused on generating hardware level programs, often hardware description language (HDL) code for field programmable gate arrays (FPGAs), from higher-level languages or design representations ~\cite{liu2016automatic, moreira2010automatic, guo2008efficient}. Although some tools~\cite{DuinoCodeGen} exist for LLM-based embedded code generation, and a small number of blog posts and tutorials explore the use of LLMs for embedded development~\cite{arduino_codegen_tutorial, arduino_codegen_blog}, these resources do not conduct a rigorous systematic evaluation of state of the art language models for embedded development and debugging or methods of interfacing directly with hardware. Instead, our work focuses on exploring and evaluating how state-of-the-art LLMs performs on end-to-end hardware-in-the-loop tasks. We emphasize the capability of these models to interface and interact dynamically with embedded systems in real-world physical environments.
\begin{figure*}[t!]
    \centerline{ \includegraphics[width=\textwidth]{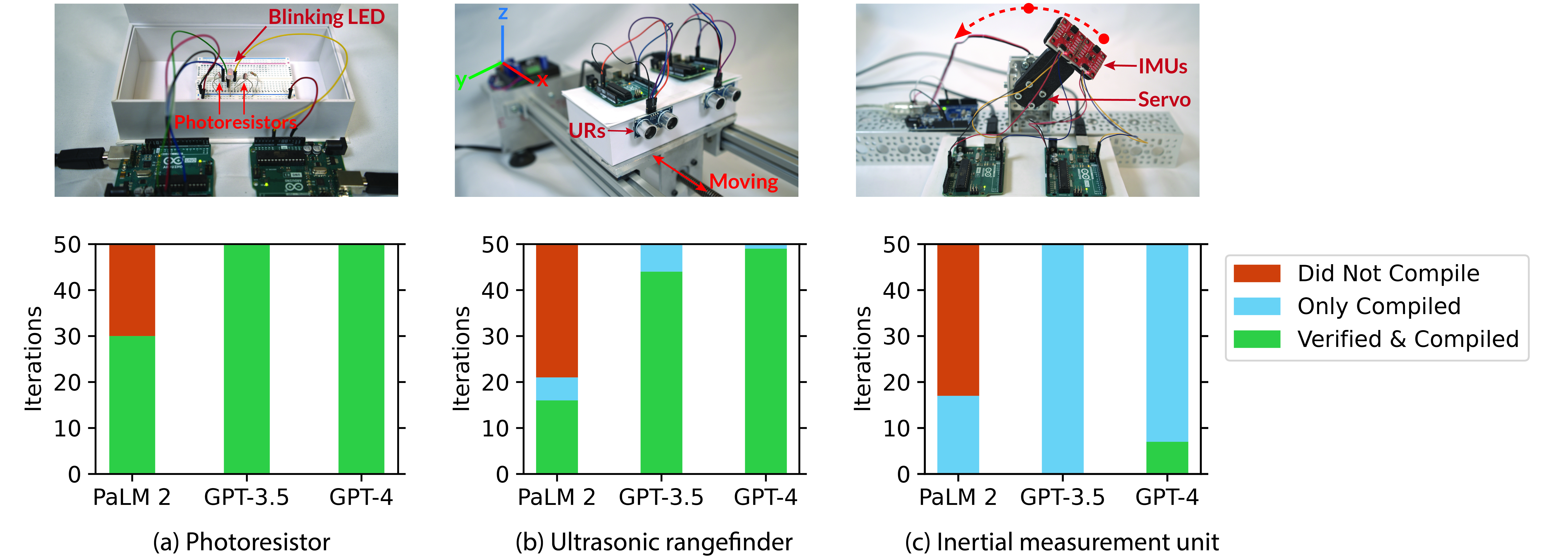}}
    \vskip -0.1in
    \caption{Evaluation results from our hardware-in-the-loop testing pipeline with $N=50$ iterations each for three tasks reading sensors on PaLM-2, GPT-3.5, and GPT-4.}
    \vskip -0.2in
    \label{fig:testbench}
\end{figure*}
\section{Physical testbenches}
To establish an evaluation framework for benchmarking LLM performance on Hardware-In-the-Loop (HIL) embedded systems development tasks, we implement an end-to-end pipeline for physical verification and fully automated, real-world testing of LLM embedded code generation. In these experiments, we specifically seek to test their performance with a single prompt and response. This is crucial because it enables programatic testing of specific embedded system tasks to identify the limits of what these models are capable of. 

To develop an interface with LLMs we leverage the key observation that given an example format within prompts, all 3 of these language models can \textit{very reliably} generate code to print data over serial. This capability allows us to create programs that produce text debugging output we can verify. We perform initial tests on a number of prompts and find this method works consistently.

To verify the code, we compare these serial outputs to outputs from another device running verified human-written programs as shown in Fig~\ref{fig:overview}. We construct physical hardware with sensor-actuator pairs described in detail below and physically attach two sensors next to each other. 

To evaluate equality, we begin by uploading the human code on both boards to understand the impact of noise and inter-device variance. We compute the Euclidean distance $d$ between the signals:
$d = \sqrt{{\sum_{i=1}^{n} (x_{2_i} - x_{1_i})^2}}$\newline
We perform N=1000 measurements and use these to determine a threshold for ``correct'' outputs (e.g. within 10\% of the reference). To align the signals we use dynamic time warping (DTW) \cite{noauthor_dynamic_2007}, a signal processing technique that corrects for time shifts, to account for variability in serial timing. 

Our fully automated testing script is compatible with the Google Vertex and OpenAI APIs and uses them to automatically prompt the model and parse the code in the response by removing explanatory text and merging code blocks into a single file. The script then attempts to compile the generated code, logs failures, and then uploads code to the MCU. It then samples both the human and LLM code output using Pyserial, and compares the signals as described above.

We run 450 experiments on three LLMs. We note automated testing is critical for fine-tuning models. This system can be extended to new devices and generalized using a logic analyzer or oscilloscope to analyze waveforms. This hardware-in-the-loop setup also presents the opportunity to extend the idea of LLMs using virtual tools and API calls~\cite{bubeck2023sparks} to \textit{physical tools}, opening up the potential for future cyberphysical systems where LLMs to query the real world through sensors. We describe our specific hardware configurations and discuss the results below. In these experiments we provide \textit{zero} additional context other than the prompt and used GPT-4 version 0314.






\subsection{Photoresistor}
Our first benchmark asks the model to read a photoresistor with the Arduino's onboard ADC.

\vspace{0.5mm}
\begin{mdframed}
\noindent\textbf{Prompt: }\textit{I have an Arduino Uno with a photoresistor attached to pin A0. Write a program that reads the value on pin attached to the photoresistor and prints it on a new line to Serial at a baud rate of 9600 every 100ms. Example serial output: 233$\backslash$n512$\backslash$n555$\backslash$n}
\end{mdframed}
\vspace{0.5mm}

We connect a photoresistor and pull-up resistor to each microcontroller and orient them toward a blinking LED inside an enclosed box without external light. Figure~\ref{fig:testbench}a shows the results of these experiments.

We observe PaLM-2 produces many compile errors, while the GPT models produce code that compiles on every iteration. The most common compilation errors result from an incorrect program format lacking \verb|setup()| and \verb|loop()| functions. We found that all programs that compiled were also fully functional and correct. This shows that LLMs can repeatably produce the correct program structure and generate functional code with a single prompt. We note the high success rate on this task is likely due to a combination of its simplicity and many similar open-source code examples.
\subsection{Ultrasonic Rangefinder}
Next we ask the LLMs to continuously read distance from an ultrasonic rangefinder. 
\vspace{0.5mm}
\begin{mdframed}
\noindent\textbf{Prompt:}\textit{Write me an Arduino Uno program for the HC-SRO4 ultrasonic ranging module that prints out the measured distance in cm without units every 100ms over Serial on a new line at baud rate 9600. The 'trig' output is connected to pin 12 and the 'echo' output is connected to pin 11. Do not use any external libraries. Example serial output: 23.273$\backslash$n23.419$\backslash$n23.366$\backslash$n}
\end{mdframed}
\vspace{0.5mm}
This sensor is more complex than a photoresistor and must be queried by setting the trigger pin high. The sensor sends an ultrasonic burst and sets the echo pin high until remains high until the sensor receives an echo. The code must read how long the echo pin is high, calculate the time difference, and then perform unit conversion using the speed of sound. 

To test the sensor, we use a lead screw driven linear actuator (Phidget Wantai Mini Stepper) to continuously move a platform back and forth across 5 cm as shown in Fig~\ref{fig:testbench}b. We mount two ultrasonic distance sensors (Sparkfun HC-SR04) and place a screen at the end of the moving platform to reflect the sensor’s transmitted waves at a consistent distance. One of the sensors is programmed with human code and the other is running on code generated by the LLM. 

Figure~\ref{fig:testbench}(b) shows the results. Again, PaLM 2 performs worse than GPT models and performance decreases overall across all models. We also observe that both GPT models have errors on some iterations. The code generated by both GPT-3.5 and GPT-4 appears to show impressive performance and correctly generates the appropriate constants to convert from time to distance based on physical constants. We note however that this particular sensor is available from numerous vendors and is also potentially well-represented in the training set. Additionally, the required constant to perform the conversion from time to distance is the speed of sound, a well-known constant that likely appears in many locations across the training dataset of these LLMs.

\subsection{6-axis IMU}
For our most complex task we test the ability of the three LLMs to generate code for a 6-axis IMU (LSM6DSO).


Our physical testbench consists of a 3D-printed servo horn that holds the two sensors side by side at an angle of 55{\textdegree} to the vertical axis. This angle ensures that when the servo (goBILDA 2000 Series Dual Mode Servo 25-3, Speed) moves through a set of positions, the IMU produces non-zero outputs on all axes. We perform verification experiments by using a third Arduino to control the servo. The testing script sends commands to move the servo to 3 angles (0, 90, and 180 degrees). At each angle the testing script logs data from the X, Y, and Z axes from LLM and user-written code for a 10s duration and checks they are within 10\%. The test script proceeds to the next angle on a correct output and only considers the code fully correct if all three tests pass.

\vspace{0.5mm}
\begin{mdframed}
\noindent\textbf{Prompt:}\textit{Write me a program for the Arduino Uno that interfaces with an LSM6DSO over I2C using only the Wire library. In my hardware implementation, the SDO/SA0 pin of the LSM6DSO is connected to GND (ground). Print the acceleration readings in gs every 100ms over Serial at baud rate 9600. Example serial output: A\_X = -0.426 $\backslash$n A\_Y = -0.023 $\backslash$n A\_Z = 0.913 $\backslash$n}
\end{mdframed}
\vspace{0.5mm}

Figure~\ref{fig:testbench}(c) shows the results. Here we observe a dramatic decrease in performance. Consistent with the prior data, PaLM 2 code rarely compiles and is not functional. For this task, we find that while GPT-3.5 consistently outputs syntactically correct programs that compile, they all fail to produce valid results. Errors include identifying the incorrect I2C address from the SDO/SA0 pin voltage (38\%) as well as hallucinating at least one of the register addresses or configuration values (100\%). In contrast, we find that GPT-4 is actually able to produce verifiable working code 16\% of the time. 

\begin{figure}[!t]
\centerline{\includegraphics[width=0.5\textwidth]{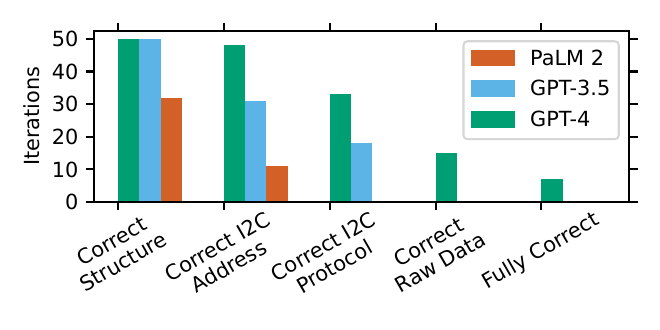}}
\vskip -0.2in
\caption{Error analysis of generated programs for the LSM6DSO IMU (N=50).}
\vskip -0.2in
\label{fig:imu_error}
\end{figure}

\noindent\textbf{Error analysis.} To probe this further, we examine the programs generated by GPT-4 to identify the types of errors made. As shown in Figure~\ref{fig:imu_error}, we observe that all of the programs generated by GPT-4 incorporate many of the elements of a correct solution, identifying the correct I2C address from the SDO/SA0 pin voltage (96\%), and of writing values to configuration registers in the setup loop and reading output registers and converting raw values to gs in the main loop (100\%), although proper configuration of and conversion from device registers proved much more challenging. An additional 16\% of programs were able to correctly read the sensor, but incorrectly scaled the resulting values to acceleration in g's. We note that this is a complex multi-step task that requires setting a configuration register to measure values in a particular range when initializing the chip and then picking the appropriate conversion constant for that mode. Additionally, the LSM6DSO was released in 2018 and development boards only recently became available through vendors like Adafruit and Sparkfun, meaning the amount of code and documentation on this sensor available before the 2021 training cutoff for GPT models is likely much smaller in comparison to the HC-SR04 ultrasonic sensor. 

These benchmarks compare the three models and suggest that although adept at simple embedded tasks, even state-of-the-art models like GPT-4 are unable to reliably produce working code from a single prompt for complex tasks, necessitating the involvement of a human developer to participate in the process. 

\section{Capabilities of LLMs for embedded systems}
The results test-bench evaluations show some capacity to develop embedded system code but with significant limitations. We build on these results and further probe the capabilities of these models to identify specific strengths and weaknesses beyond single-shot code generation. 

\subsection{Hardware Specifications}
We find that GPT-4 possesses a surprising amount of knowledge of specific integrated circuits, down to individual register addresses and operating modes. For example, GPT-4 correctly identifies the two I2C addresses for the Bosch BME280 and explains they are dependent on the voltage applied to the SDO pin. In addition, it produces a list of common control registers for interfacing with the chip. To further confirm this, we follow the same procedure with the AD5933, an obscure complex impedance measurement chip with few online examples.

While descriptions of some registers are self-explanatory, such as the MSB and LSB of sensor readings, the function of registers with multiple control or flag bits is ambiguous. Upon further prompting, GPT-4 clarifies what these individual bits represent. We note that some of its explanations are direct copies of the datasheet, suggesting that this understanding of hardware components is at least partially due to datasheets being included in the training dataset. 

We find that overall performance on correctly identifying hardware specifications is mixed, GPT-4 sometimes incorrectly identifies the size of each bit field within a register. Additionally, GPT-4 correctly identifies the normal operating voltage range for the BME280, but incorrectly identifies the absolute maximum ratings. It is also not aware of all integrated circuits, especially those that are less frequently used or referenced online. For example, we found it was consistently unable to identify the I2C address of the MAX9611 amplifier. 

\subsection{Software}
It is well-understood in the existing literature that LLMs can generate code in various programming languages. We find that GPT-4 can utilize the embedded understanding of hardware to generate code and approaches that appropriately utilize hardware components. When developing embedded system code, one particular pain point is generating a high-level program structure. One useful application of LLMs is generating driver and program templates. For example, we prompt GPT for a template of an IMU driver, and GPT provides a template along with instructions on the specific actions that need to be taken in each step, such as defining constants, specifying the device I2C address, creating appropriately scoped output variables, and initializing and reading the chip. This could be particularly useful for inexperienced developers in guiding them to develop readable and well-organized embedded code in addition to improving the efficiency of those with experience. 

Additionally, models can apply the understanding of IC register addresses and communication protocols like I2C/TWI to generate working code for basic interface functions, such as ``get'' and ``set'' methods for reading or writing to sensors and actuators. With no additional context, this only works well for simple components which the model also demonstrates a strong understanding of the underlying register addresses and functions. However, performance in this area can be vastly improved by first prompting the model with a machine-readable representation of important registers and functions from the component datasheet or providing an API or header file from an external library or driver. 

This is particularly powerful when combined with generated templates, enabling programmers to leverage GPT-4 to develop high-level outlines and follow up by implementing modular portions of the code. We verify this approach by prompting GPT-4 with a register table for a fictional temperature sensor IC (guaranteed not to be in the training data) and requesting pseudocode for a basic driver for this chip, and find GPT-4 produces code consistent with the register table and descriptions as shown in Figure ~\ref{fig:fake_chip_code} (see Appendix for full register table). 
\begin{figure}
    \centering
    \includegraphics[width=0.6\textwidth]{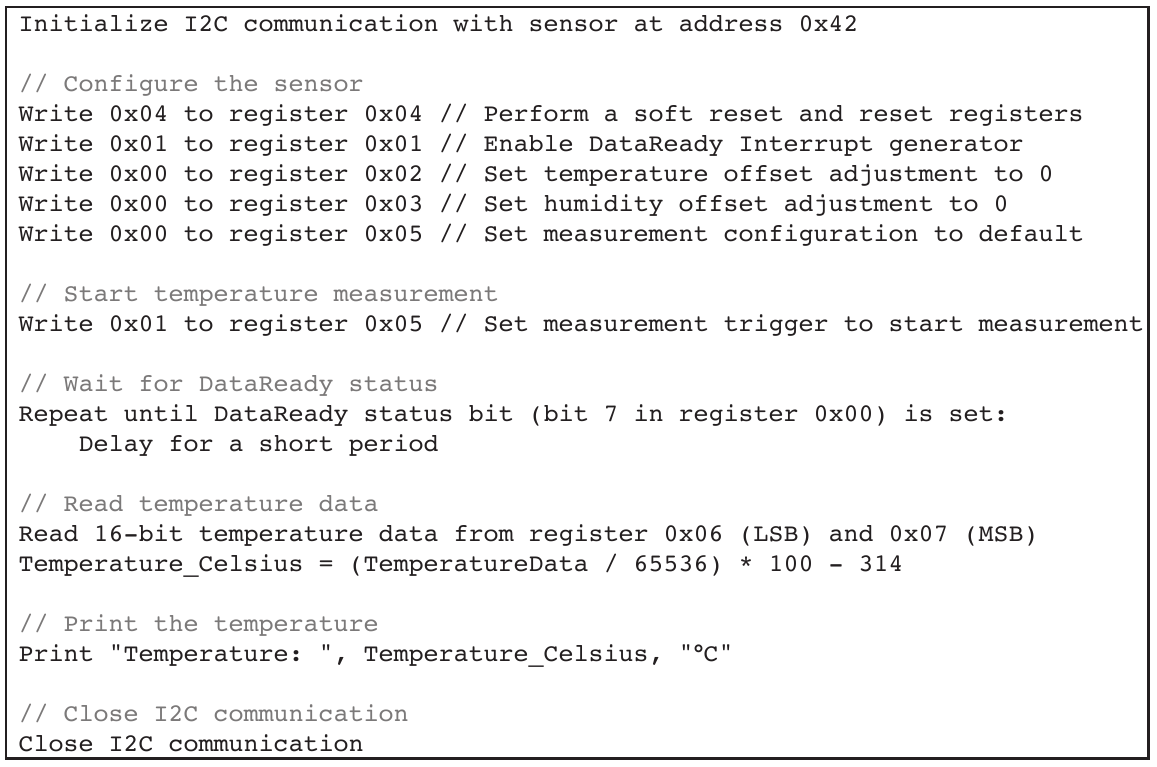}
    \caption{Pseudocode generated by GPT-4 to read the temperature from a fictional environmental sensor IC based on a provided register table.}
    \label{fig:fake_chip_code}
\end{figure}
\subsection{Design} 

In embedded systems, a significant part of the work involves designing hardware and software setups based on real-world signals. Even when programs become too large or complex for these LLMs to reliably generate working code, we find that it is still able to provide potentially useful suggestions for designers working to build or optimize a system, including hardware platforms, communication protocols, and code-writing techniques using relevant libraries. We observed similar behavior in numerous interactions throughout our testing and user study in which the models offer design and debugging advice.

To further probe this, we ask LLMs to provide optimization suggestions for a C program built with the Nordic nRF5 SDK that uses an nRF52832 microcontroller and HDC2010 sensor to advertise the temperature every 250 ms. Our initial program consumes over 9 mA of current and has no power optimizations. To solicit optimization suggestions, we prompt each model with the text below.

\begin{figure}[!t]
  \centering
  \subcaptionbox{LLM optimization suggestion frequency.\label{fig:probability_power_optimization}}{%
    \includegraphics[width=.4\textwidth]{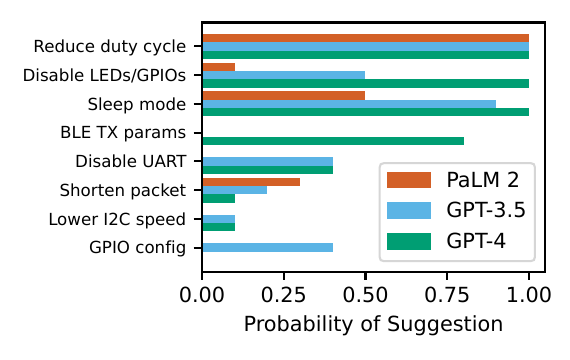}
  }
  \subcaptionbox{Current reduction for each optimization\label{fig:power_savings}}{%
    \includegraphics[width=.4\textwidth]{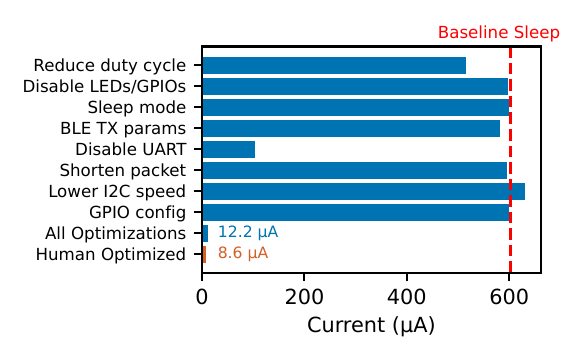}
  }
  \caption{Probability of the models suggesting each power optimization in repeated trials (left). Power savings from LLM suggested optimizations. The original code consumes 9.1 mA. Combining all LLM suggestions lowers the current to 12.2 \textmu A, slightly higher than the human expert's 8.6 \textmu A (right).}
  \label{a-two-column-figure}
\end{figure}

\vspace{0.5mm}
\begin{mdframed}
\noindent\textbf{Prompt: }\textit{We are trying to design a battery-operated environmental sensor built around a nRF52832 and an HDC2010 that transmits temperature measurements using Bluetooth advertising packets to a receiver. We will use this data to monitor the change in ambient temperature inside a building over a month. This code already reads the sensor and transmits data correctly. Please suggest the specific parts of this code we could modify to maximize battery life: **program**}
\end{mdframed}
\vspace{0.5mm}
We repeat this prompting for ten iterations with each model, and record the occurrence of specific and actionable improvements. For example, a response that states \textit{``When not reading from the sensor or transmitting data, make the device go into sleep or power down mode... in the main loop, you can put the device to sleep instead of simply running an idle process.''} would count as a sleep mode suggestion while a vague response like \textit{``check if your microcontroller has a sleep mode''} would not count. We categorize the frequency of each of the optimizations in ~\ref{fig:probability_power_optimization}.

We observe that despite not having the word ``power'' in the prompt, all the models inferred this from the mention of battery life. Enabling sleep mode had the most significant impact, reducing power consumption to 602~\textmu A, and we then implement and benchmark the effect of each individual suggestion relative to this baseline. As shown in Figure~\ref{fig:power_savings}, we find that all suggestions, except for lowering I2C speed, resulted in decreased power consumption. By incorporating the models' power consumption strategies, the system consumed an average of 12.2 \textmu A, similar to the optimization by an experienced developer (8.6 \textmu A). The improvements between the LLM-generated suggestions and human optimization include correctly recognizing that decreasing the I2C speed will hurt power consumption and enabling the DC voltage converter on the nRF52832. 

This experiment demonstrates the ability of these models to generate reasoning across hardware and software domains and identify how code implementation affects the real-world performance of a system. Additionally, the models can produce reasoning about the signals and factors that affect power, such as consistently suggesting a lower sampling rate.  Although no single prompt identified all of these power-saving improvements, we believe performing a process like this could be extremely beneficial to designers when optimizing their system performance. While an inexperienced developer might simply implement the suggestions provided by a LLM and achieve a promising result, we again find that human evaluation is critical, since one suggestion, decreasing i2c speed, actually slightly increased the power draw of the system. 

\section{Limitations of LLMs for Embedded Systems Development}
We categorize the limitations of LLMs into two categories: general limitations of their programming and reasoning abilities that appear in other studies~\cite{chen2021evaluating}, and limitations specific to developing embedded systems. With embedded systems, representation is a significant challenge due to the added physical components and data sheet referencing paradigm used throughout embedded systems development.

\subsection{Task understanding}
LLMs often incorrectly interpret user intent, typically by making incorrect assumptions based on unintentional ambiguity in prompting \cite{aina-linzen-2021-language}. As part of our user study, participants were tasked with reading a PPG sensor and transmitting the analog values from the sensor over Bluetooth to a receiver in order to view the pulse waveform in the serial plotter. Many participants described the task to ChatGPT as \textit{"reading an analog heart rate sensor and transmitting the values."} GPT-4 regularly interpreted this to mean the analog signal was the heart rate itself as opposed to a raw PPG signal, and provided code to transmit this value periodically as a wireless heart rate monitor instead of streaming PPG data. While a human programmer working physically with these sensors may understand based on the sensor type what is desired, LLMs often subtly misinterpret tasks like this and can produce code that compiles and runs, but does something different than what the user intended. 

\subsection{Hallucinations}
Another well-documented limitation of LLMs is hallucination where models produce incorrect or misleading details in a response. In the context of embedded systems, we see this issue arise with models generating nonexistent pin numbers, register names, and library references. This may also result from a lack of system-level context, as LLMs frequently make assumptions for unknown or ambiguous values rather than expressing uncertainty \cite{aina-linzen-2021-language}. Expert understanding of the embedded system may alleviate this issue by allowing the user to better prompt the model or debug the output. 


\subsection{Unprompted behavior}
We find in some cases that LLMs will make unprompted modifications to code. For example, the model may add unnecessary print statements, causing issues in resource limited systems. Similarly, when queried to generate code and then requested to make changes to that same code, LLMs may make additional unnecessary and unprompted changes along with the requested updates which can easily go unnoticed. This may be remedied through an understanding of the embedded software, by prompting the model to only make changes to specific lines or sections, or future automated tools to check for these errors.


\subsection{Version and Library errors}
We find another common error made by LLMs is confusing different open-source libraries and their versions. In one instance, when asked to generate code for the LSM6DS3 IMU, the resulting output contained calls to different open-source libraries from Adafruit and Sparkfun. While each of the libraries exists and the function calls to them were valid, instantiating an object with one library and trying to call methods from the other will obviously produce errors. This highlights a difference in the types of errors made by LLMs as compared to human programmers. Similarly, we find that in some cases LLMs may use older or deprecated function calls. While this may sometimes produce compile errors, it is possible for these programs to compile and run while exhibiting more subtle errors. In contrast with mixing libraries, these errors are similar to those made by human programmers. This also suggests a path for future automation and improvement where an AI enabled IDE could verify the compiled outputs and feed the errors back into the model to correct them


\section{AI based embedded development workflow}

\begin{figure}[!t]
\centerline{ \includegraphics[width=0.7\textwidth]{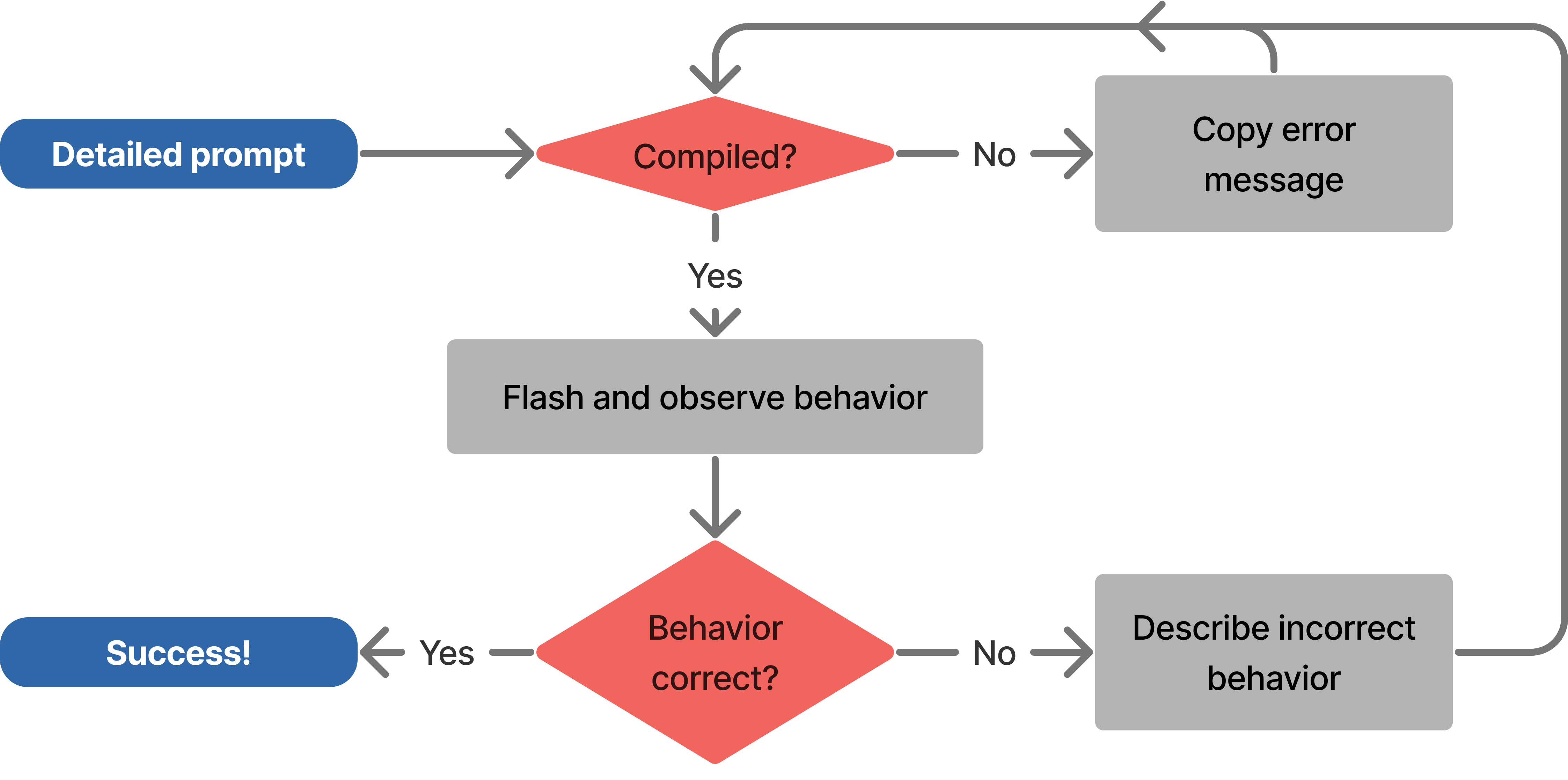}}
\vskip -0.2in
\caption{Flowchart with our proposed high-level workflow using LLMs for embedded system development.}
\vskip -0.2in
\label{fig:workflow}
\end{figure}

In this section we synthesize our learnings from the experiments conducted above and propose a workflow for integrating LLMs into embedded systems development. A suggested workflow is visualized in Figure~\ref{fig:workflow}.

\subsection{Providing Detailed Prompts}
Prompt engineering to provide LLMs with ample context is important to generating useful code. Prompts should provide a complete scope of the task including relevant details about both hardware and software. For example, sample code snippets, specific libraries, and chip versions (either part number or specific development board, e.g. ``Arduino Uno'') can all improve performance. Additionally, in multi-device systems with physical peripherals, it is important to specify part numbers for both devices and explain the functions and interactions they perform or provide example code. We provide examples below.

\vspace{0.5mm}
\begin{mdframed}
\noindent\textbf{Bad prompt:}\noindent\textit{``Make the LED blink and print hello world.''}
\end{mdframed}
\vspace{0.5mm}

This prompt does not give a complete specification of the desired behavior such as the specific microcontroller to use, or the LED blink rate, etc. While this may seem obvious to a human programmer working with a specific development board, this context is explicitly needed by the model. Providing such context could be easily automated in future IDEs where target hardware must be specified for compilation.

\vspace{0.5mm}
\begin{mdframed}
\noindent\textbf{Better prompt:}\noindent\textit{``I have an Arduino Uno with an LED attached to Pin 13. I want to print the message ``Hello World'' to serial at baud rate 9600 and blink the LED on and off every second. First explain how to approach the task and then provide a program to accomplish it.''}
\end{mdframed}
\vspace{0.5mm}
In comparison, the above prompt gives a complete specification of the problem and the desired implementation.

\subsection{Debugging Compilation Errors}
The first step in validating the generated code is to attempt to compile it. Compilation may fail for a number of reasons and is a means to catch a number of common errors discussed above such as hallucinations of nonexistent libraries, peripherals, or pins. This process is familiar to expert programmers who often face similar issues when trying to adapt code examples to their target application. We find however that LLMs are also capable of automatically correcting their mistakes when error messages are copied back into the model as new prompts.

\vspace{0.5mm}
\begin{mdframed}
\noindent\textbf{Prompt:}\noindent\textit{“When I try to compile this code, I get the following error: ***copied error here***”}
\end{mdframed}
\vspace{0.5mm}

Often, compiler errors are difficult to parse and relate back to the code, particularly for languages like C/C++ where the source of the error may not be in the line explicitly referenced by the compiler. We find that GPT-4 is regularly able to relate these error messages back to the generated code to identify and resolve issues like incorrectly used libraries and syntax errors when provided with the compiler error message. 

Additional context can be provided as well, for example if the error message is related to the use of a library that is not available for a specific platform, suggesting an alternate library in addition to providing the error message can help steer the language model towards resolving the issue. Similar to the compiler context above, this step could be automated by feeding the compiler output directly into the LLM to correct its errors or to fine-tune models for correcting errors. 

We note however that an automated system may perform better with a limited number of ``retries''; we find that when models fail multiple times to fix an issue, it is often best to start over with a new chat. This occurs as LLMs produce outputs dependent on past user inputs and model outputs (referred to as tokens). 

These token windows are finite, meaning continuous incorrect outputs from a model may clutter the input sequence used by the model resulting in error-prone outputs. Additionally, users can provide additional information at the start of a new chat learned from previous failed chats, for example explicitly specifying not to use a specific library that the language model struggled to implement in a previous chat. 

\subsection{Program and Behavior Observation}
After verifying that the code is capable of compiling and uploading, the next step is to test the behavior of the code itself. For simple tasks like blinking an LED, this may be verified visually, however comprehensive testing and verification of more complex embedded systems is non-trivial and will likely continue to require a human in the loop. For example verifying the output of a sensor or measuring power consumption requires physical tools. We note however that these processes could be further automated through the development of physical unit testing platforms as described above that can provide automated feedback and virtual interfaces to physical tools. 

\subsection{Describing Incorrect Behavior}
Perhaps the most surprising result of our experiments is that LLMs, GPT-4 in particular, can provide detailed and useful advice for helping debug embedded systems code. Similar to the initial prompt, we observe the best results when providing the model with a detailed explanation of what exactly is wrong with the behavior of the system and ask it to fix the code. We provide example prompts below.

\vspace{0.5mm}
\begin{mdframed}
\noindent\textbf{Bad prompt:}\noindent\textit{“This doesn’t work”}
\end{mdframed}
\vspace{0.5mm}

We find in some cases simply telling the model the code does not work or providing other negative feedback like ``no'' can prompt the model to recognize mistakes like syntax errors, but this is less effective for more subtle issues concerning system behavior. Generally a more specific prompt that details precisely how the behavior differs from what is desired will yield more relevant responses decreasing the number of queries required. In particular, we find that GPT-4 is often able to identify relationships between descriptions of a system's physical behavior and the associated code and identify sources of errors based on these physical description. 

\vspace{0.5mm}
\begin{mdframed}
\noindent\textbf{Better prompt:}\noindent\textit{“This code compiles and runs and the Arduino prints messages, but these messages have incorrect zero values for the A\_X, A\_Y, and A\_Z acceleration channels instead of acceleration readings. The current output we are getting is: A\_X = 0, A\_Y = 0, A\_Z = 0''}
\end{mdframed}
\vspace{0.5mm}

In comparison, this prompt tells the model both what works and the specific issue that needs to be fixed, in this case the zero values for each of the acceleration channels. In this instance, GPT-4 correctly identified that the issue was likely due to an incorrect i2c configuration and provided steps to debug, first advising to check the physical connections of the power, ground, signal, and clock lines, and then providing a version of the code with the I2C address changed from \verb|0x6A| to \verb|0x6B|, which resolved the issue. 
We note here that perhaps the most surprising results of our study were that models such as GPT-4 are able to provide advice for debugging potential hardware issues that require an understanding of the physical system. For example, the model would ask the user to check physical connections or check wiring. We observed this multiple times in our user study discussed below. 

\subsection{Additional Tips}
\noindent\textbf{GPT often struggles with numbers.} This is commonly observed in other applications \cite{chen2023chatgpts}, and we find this specific application space is no different. It can often be more efficient to manually fix numerical and calculation errors in the code instead of prompting the model to fix them. 

\noindent\textbf{Human in the loop.} Remember that GPT-4 and other models in their current state should be viewed as \textit{tools to aid human users} in programming and not one-shot solutions. If you find a bug that you know how to fix, fix it yourself. For example, if you ask the model to flash an LED on Pin 13, but it responds with a program that flashes Pin 12, manually updating the pin number is much more efficient. Similarly, providing as detailed an explanation as possible about the error can significantly improve the model's ability to help debug and solve the problem as shown in the prompt below.

\vspace{0.5mm}
\begin{mdframed}
\noindent\textbf{Prompt:}\textit{“I asked to use the Sparkfun IMU library for this, but I see there are some calls to the Adafruit IMU library in the setup() function. Please rewrite this to use only the Sparkfun library to interface with the IMU.”}
\end{mdframed}
\vspace{0.5mm}

This prompt shows an example of correcting the mixed libraries error noted above. While the actual error behavior resulting from using different libraries may be complex, providing this observation of the error in the code to the model can help guide it to a correct response. 

\noindent\textbf{Keep the context up-to-date.} Adding context of any user modifications to the generated code is important. For example, after correcting errors as discussed above or editing the code, the new version should be copied back as part of the prompt before generating additional code. We find that the exact format is not important. For longer programs, copying in the segments of updated code along with a brief natural language description of them is sufficient to provide the model context to integrate it into new solutions.

\noindent\textbf{Give examples.} One powerful technique is to create prompts with example code and explain what you'd like the model to change. Just like a human programmer, LLMs can often be more successful when given an example code that is close to the target solution. For example, when asking for a receiver, provide the transmitter code. Similarly, providing an API to constrain the output can also help.

\noindent\textbf{Restarting.}
If GPT fails multiple times to fix an issue, it is often best to start over with a new chat. This strategy is helpful for multiple reasons. First, models such as GPT-4 have limited context windows. In a long chat where responses include blocks of code consisting of many tokens, the initial prompt may fall outside this window causing the model to lose reference to the original goal. Sometimes subsequent programs and chats are enough to maintain this information but for example, a long chat about debugging hardware connections could lose the initial discussion about the code. Second, all of the previous responses and chats are used to generate the next response. If the model's initial suggestion had significant errors, future responses will still be conditioned on this content if the user is unable to precisely articulate what is wrong with those responses.

\noindent\textbf{Be explicit about changes.}
We note that it is important to be explicit about what should be changed and what should be kept the same to prevent the model from making unprompted changes as shown in the following example prompt.

\vspace{0.5mm}
\begin{mdframed}
\noindent\textbf{Prompt:}\textit{“I have a program for the Arduino Uno that reads data from an LSM6DSO IMU using the Adafruit IMU library and prints the data over serial. Instead of printing to serial, I want to analyze this data and flash an LED on Pin 13 every time the total magnitude of the acceleration is higher than 2 Gs. Please replace the portion of code that is printing data to serial with code to do this calculation, leaving as much of the original program unchanged as possible.”}
\end{mdframed}
\vspace{0.5mm}

\noindent\textbf{Break down tasks.}
We observe that LLMs are best at generating short to medium-length programs. If the model fails multiple times to generate complete working code, try asking it to develop modular functions individually and integrate them into a complete program yourself. This is similar to the chain of thought strategies proposed in \cite{wei2022chain}. Another strategy is to begin by asking the model to generate pseudocode for a particular task, and then asking it to complete those functions and combine them into a complete program.

\section{User Study}

\begin{figure*}[ht!]
    \includegraphics[width=\linewidth]{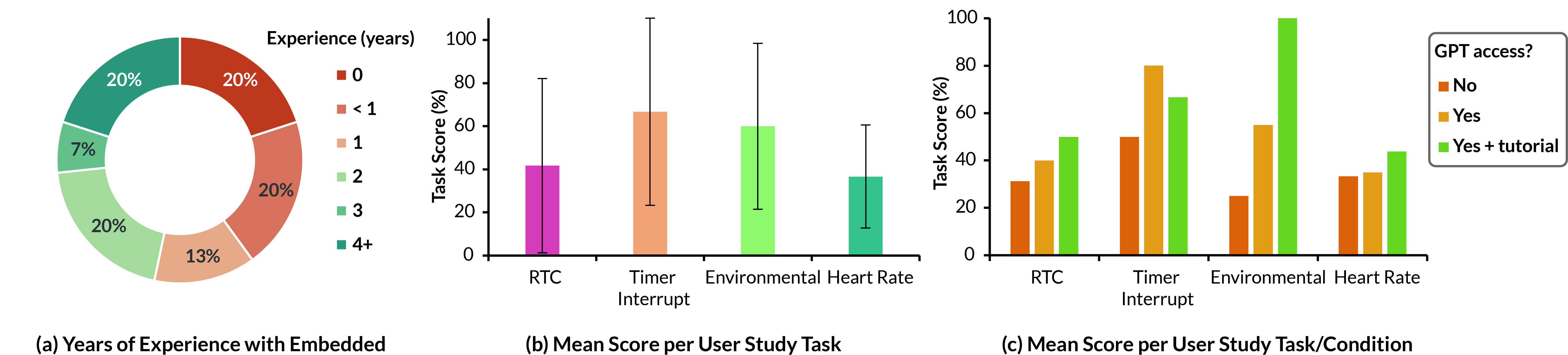}
    \vskip -0.1in
    \caption{Left: Distribution of user study participants' experience with embedded systems. Middle: Mean scores obtained by participants on each task. Right: Mean scores obtained by participants on each task split by condition.}
    \label{fig:user_study}
\end{figure*}

We perform a user study to evaluate our workflow guidelines and investigate the ability of LLMs to aid in hardware debugging.

\subsection{Methods}


\noindent\textbf{Participants.}~We recruit 15 users (age $M=23.3$~years, $SD=2.93$~years) comprised of undergraduates and PhD students, faculty, and professional software engineers.
Although all participants had an electrical engineering or computer science background, their experience with embedded systems experience was evenly distributed from 0 to 4+ years as seen in Fig.~\ref{fig:user_study}.

\noindent\textbf{Setup.}~Studies were conducted in a quiet lab environment.
Participants were given a laptop with the Arduino IDE and all necessary device drivers installed.
A web browser with ChatGPT using the GPT-4 0314 model was open by default, and participants were told they could use a search engine of their choice.
Participants were also provided a kit of components with microcontrollers, sensors, actuators, wireless transceivers and jumper wires.
The specific components are enumerated as part of the procedure.

\noindent\textbf{Procedure.}~We ask participants demographic questions to gauge their experience with programming, C/C++, and embedded systems.
Participants were assigned a unique ID number, used to select a row in a counter-balanced partial Latin square to determine the order of the tasks and conditions.
Each participant completed a simple warmup task and four graded tasks.
We recorded the laptop screen, a top-down video of participants' interactions with the hardware components, their chat history with ChatGPT, and their final submitted code.
After completion, we conducted semi-structured interviews with quantitative questions based on the NASA Task Load Index (TLX)~\cite{hart1986nasa} and SPACE Framework~\cite{forsgren2021space}. 
Participants were encouraged to elaborate on their responses to the quantitative questionnaires. 

\noindent\textbf{Study Design.}~The study was a $5x2$ between-subjects design. 
Participants were timed and completed all four tasks in a random order and with varying levels of GPT-4: no access, access to GPT-4, and access to GPT-4 with our workflow guidelines.
We used a subset of the Latin square such that participants never had access to GPT-4 with our workflow guidelines before having access to GPT-4 alone.
We designed the tasks to be similar to assignments from an introduction to embedded systems class, and they included:
\begin{enumerate}[leftmargin=*]
    \item \textbf{Warm-Up} (untimed, $\sim$5 minutes): Print ``Hello, World!'' to the Serial console and blink an LED.
    \item \textbf{Environmental Sensor} (40 minutes): Connect the BME280 environmental sensor to the Arduino Uno (with already attached RFM9x LoRa radio) and transmit sensor readings to a second Arduino Uno with LoRa radio.
    \item \textbf{Heart Rate Monitor} (40 minutes): Connect the analog heart rate sensor to the nRF52 (Arduino Nano 33 BLE) and transmit sensor readings to a second nRF52 over Bluetooth Low Energy.
    \item \textbf{Timer Interrupt} (20 minutes): Connect the piezoelectric buzzer to the Arduino Uno and generate a tone of 440 Hz using built-in hardware timers.
    \item \textbf{RTC Wakeup} (20 minutes): Configure a PCF8523 RTC (real-time clock) to wake an Arduino Uno from sleep mode to flash the LED before going back to sleep.
\end{enumerate}


\noindent\textbf{Analysis.}~The research team reviewed the submitted code and graded it according to a rubric, with partial credit awarded for each sub-task successfully completed.
This rubric was not provided to participants.
We summarized the participants' experience with the study through descriptive statistics derived from these scores and participants' responses to the survey instruments, using a Student's t-test to determine significance in these results.
In addition, the interview transcripts were reviewed for thematic analysis.
Quotes from the interviews and behaviors observed from the video recordings were extracted to substantiate those themes.

\subsection{Task Performance Results}

Participants scored highest on the Timer Interrupt task ($M=66.67\%$, $SD=43.46$), and worst on the Heart Rate Monitor task (score $M=36.67\%$, $SD=23.92$); the variance suggests the Heart Rate Monitor task was difficult across experience levels, while the Timer task was easier for experienced users.
Indeed, the Heart Rate Monitor task required multiple components that introduced multiple points of potential failure. P10 cited how this complexity made it challenging to work with GPT: \textit{``The first task [Bluetooth]... I'm not sure if it would have been easier with GPT or not... I had trouble like with everything on that... it was hard to explain the system... that you're building here to GPT.''}

Participants performed the best with access to GPT and our workflow, significantly better than with no access to GPT (mean score 65.48 vs 34.21, $t(38) = 2.73$, $p << 0.01$), and better than using only access to GPT with no significance, as seen in Figure~\ref{fig:user_study}c. 
In the Environmental Sensor task all 5 participants with GPT access and our workflow scored 100\%, and half had little to no embedded systems experience.
Participants without GPT scored 25\% on average. We attribute much of this success to our tutorial.
For example, using clues from P7's detailed description of the Arduino's output, ChatGPT explained Arduino's implementation of \verb|sprintf()| does not support formatting floats. 
ChatGPT then corrected its own code, using \verb|dtostrf()|, enabling P7 to successfully complete the task when otherwise \textit{``I don't think I would have been able to do the LoRa sensor in 40 min.''}

\subsection{Interaction Observations}

We also observed interesting interactions with GPT during these studies:
\begin{enumerate}[wide, labelwidth=!, labelindent=0pt]
\item \textbf{Many participants commented on how GPT was useful for generating starter code.}
Similar to an Internet search for examples as P9 put it, \textit{``Instead of searching for example code and then modifying that... you know it'll generate this code [that's] more customized to what you want.''}
From an efficiency standpoint, P6 suggested that, \textit{``...the one good thing about ChatGPT -- because when you search on google there are a lot of things you cannot use, and you have to click on them one by one, and you don't know what you want -- maybe he [chatgpt] provides failing code, but you know it\'s what you want, and you can modify it based on that.''}
The ability to get programmers started, and especially those that are inexperienced, with bootstrapped code is extremely powerful and motivating.

\item \textbf{Experienced programmers asked GPT for smaller chunks of code.}
For example, a function to read a sensor, a function to send wireless transmissions, and then synthesized them together.
P7, an experienced programmer with \textit{zero} hardware experience, provided this context: \textit{``Hello Chatgpt I want help writing code for a big task, we will break it down to two tasks. Let's first do the first task then we will do the next task which I will ask you to incorporate the first task along with the results of the second. Does this sound good? ... so our first task will be to read the pressure, temperature and humidity data from our BME280 sensor which is connected to our arduino uno board... great the first task is done and we have correctly got the output we wanted now lets start the second task which we will use the first task to get started. For this second task we are going to have another arduino uno board which has a LoRa radio RFM9x which has the corrasponding ports...''}
By working with smaller tasks, GPT was more able to effectively debug and iterate.

\item \textbf{ChatGPT offered specific and actionable hardware debugging advice.}
This included checking pin connections and wiring and was helpful to all participants, especially those with no hardware experience that struggled to interpret advice found online: \textit{``I saw a lot of the words [online] like, you have to connect to a ground I didn't really know what that meant... I figured I'm assuming that the colors of the cables don't mean anything''}~(P7).
\end{enumerate}
In one particular case, ChatGPT went beyond the suggestion to simply check the connections and even provided a custom debugging script to scan for I2C addresses and verify that the sensor could be found.
\vspace{0.5mm}
\vspace{0.5mm}
\begin{mdframed}
\noindent\textbf{Person:}\noindent\textit{``I tried to run the code with both 0x76 and 0x77, but both print "Could not find a valid BME280 sensor, check wiring!"''}

\noindent\textbf{ChatGPT:}\noindent\textit{``If your Arduino says it cannot find a valid BME280 sensor, it might be due to one of the following reasons:}

\begin{enumerate}[leftmargin=*]
\item \textit{Wiring issue:} \textit{Double-check your wiring connections as explained previously. Make sure the SDA and SCL connections are correct. The SDA pin should connect to A4 and the SCL pin to A5 on the Arduino Uno. Also, ensure the sensor is correctly powered with 3.3V and its ground is connected to the Arduino's ground.}
\item \textit{Incorrect I2C address:} \textit{The I2C address of the sensor can typically be 0x76 or 0x77. If neither of these addresses is working, your sensor might have a different address. You can use an "I2C scanner" sketch to find out the exact address of your sensor. Here is an example of an I2C scanner code: [I2C scanner program]''}
\end{enumerate}

\end{mdframed}
\vspace{0.5mm}

\subsection{Exit Interview Feedback}

Finally, we evaluate participants' responses to our subjective questions.
Through responses to the TLX questionnaire, we learned about participants' perceived load of each task depending on their GPT access condition.
The average perceived load for conditions in which participants had access to GPT and our workflow guidelines was lower than the perceived load when participants had access to GPT only (mean load 3.25 vs 3.83, $t(38) = 1.49$, $p ~= 0.1$), and also lower than the perceived load when participants had no access to GPT (mean load 3.25 vs 4.67, $t(35) = 4.35$, $p << 0.01$).
Responses to the SPACE framework questionnaire also sparked interesting open-ended discussions from participants.
Participants were largely split on every aspect: some responses suggested that ChatGPT has the potential to help you learn, while others hinted that it could hinder learning.
P4 did not feel very fulfilled \textit{``...because, like, I feel like I didn't [do any] work. You know, like it wasn't something that I did. It's something that ChatGPT did for me... it was cool to see that it was working, but not nearly as cool as if I had done it personally.``} On the other hand, P8 pointed out that \textit{``...it's fulfilling in different ways: ...instead of understanding the domain, you become more fulfilled in that you get the task to work... GPT handles some tedious things really well, and allows you to better understand some of the other aspects of the task.''}


\section{Discussion and Conclusion}
We demonstrate LLMs have capabilities far beyond code auto-complete utilities - they show a promising ability to work across hardware and software domains as well as rationalizing how code effects the \textit{physical behavior} of real-world systems. We summarize key takeaways below and outline directions for future work.

\vskip 0.05in\noindent\textbf{LLMs enhance human developer productivity.} While LLMs cannot reliably produce working end-to-end systems on their own, we conclude that these models can significantly increase productivity and satisfaction by providing automation and a level of abstraction for experienced developers who provide context-rich prompts. These tools can also empower novice developers as shown in our user study and consistently assisted users in debugging across both hardware and software. In particular, GPT-4 provided useful hardware related debugging steps and identified sources of incorrect behavior (e.g. incorrect physical connections, packet structure, or I2C addresses) for errors not caught at compile time. These results highlight the potential of LLMs to dramatically reduce the high barrier to entry for working with hardware by providing problem specific debugging advice. For example, we observe through teaching experiences that many students in embedded systems courses with only software experience struggle to apply their programming skills when hardware is involved. Future work to develop these findings into a teaching tool has significant potential to increase accessibility to physical computing and fabrication research within the Ubicomp community.

\vskip 0.05in\noindent\textbf{Prompting is critical.} We discover through our exploration of LLM capabilities and confirm through our user study that the development of effective prompts is crucial. Prompts must unambiguously encapsulate key system information to enable LLMs appropriately engage with the technical nuances of both the physical hardware and the intended behaviors. Although we propose initial guidelines for users, there remains much room for future work in rigorously identifying the most effective methods for optimal collaboration between human developers and LLMs on embedded development tasks. Performance in this area could also be improved through strategies such as fine-tuning Large Language Models (LLMs) on specialized embedded repositories, datasets, and specific Software Development Kits (SDKs). Additionally, developing new ways of efficiently encoding the full system state to provide context to the model could further improve performance. We see our framework for automated hardware-in-the-loop testing that would allow the models to query the system state like a software plugin as a powerful tool to support future work in these areas. 

\vskip 0.05in\noindent\textbf{Benefits of hardware-in-the-loop.} In this work, we consider the case of a human developer interacting with an LLM through a text interface. However, there are other interface techniques that could be explored. By integrating LLMs into future Integrated Development Environments (IDEs) and leveraging hardware-in-the-loop test benches, LLMs could programmatically overcome the limitations highlighted earlier. This integration could allow for automated and iterative design enhancements in real-time during development. By harnessing recent advancements in visual \cite{zhang2023videollama} and sensor \cite{girdhar2023imagebind} embeddings, upcoming research could unveil methods to represent hardware configurations and setups through consistent and unambiguous embeddings. Such developments would enable language models to generate more informed software solutions grounded in specific hardware requirements.

\vskip 0.05in\noindent\textbf{Cyber-physical AI systems.}
Our observations demonstrate the potential to create a new class of cyber-physical embedded systems powered by LLM-based autonomous agents \cite{wang2023survey, xi2023rise, sumers2023cognitive, yao2022react}. While such models are too computationally intensive to run locally, internet connected IoT devices could interact with AI agents directly or via an intermediate edge device. We expect future AI agents will play key roles in the observation, planning, and action stages of ubiquitous computing devices. In the observation stage of an environmental sensor, agents could detect anomalies to identify physical sensor faults or react to changes in the environment and diagnose performance issues in real-time. In the planning stage these insights could be used to generate a self-correcting software update to reset or disable a broken peripheral, or adjust communication parameters to account for changes in a wireless channel. In the action stage, the agent will implement these solutions by deploying code adjustments or re-calibrating system parameters in real time to maintain system integrity or functionality. Such closed loop systems would enable LLM-powered agents to not only monitor and react to system behaviors but also proactively manage and enhance embedded system operations for self-regulated continuous improvement. 

\vskip 0.05in\noindent\textbf{Role of humans and responsible development.}
Importantly, our user-study feedback regarding fulfillment, though generally positive, was at times ambiguous. Some participants expressed that LLM assistance made them feel less responsible for the resulting systems. As the capability of language models and tools that leverage them continue to grow, it will be critical to define the role of human developers in the process. This is particularly important in cases where embedded systems are used for mission-critical subsystems, where ownership and extensive verification of code can have significant safety implications. Large language models have the potential to re-imagine how embedded system development and debugging function, but creating direct interfaces between these systems and the physical world simultaneously introduce profound risks. It is imperative that future work proceeds responsibly to address these concerns and does not lose sight of our responsibility, as embedded systems researchers, to build ethical and safe systems.

\bibliographystyle{ACM-Reference-Format}

\appendix
\clearpage

\begin{table}[]
\begin{tabular}{p{0.2\linewidth}p{0.2\linewidth}p{0.2\linewidth}p{0.2\linewidth}p{0.2\linewidth}}
\multicolumn{5}{l}{\textbf{Appendix: Register Map for ABCD500 \noindent Temperature and Humidity Sensor}} \\
\multicolumn{5}{l}{I2C address 0x42} \\
ADDRESS (HEX) & NAME                               & RESET VALUE               & DESCRIPTION                            &  \\
0x00          & INTERRUPT/DRDY                     & 00000000                  & DataReady and interrupt configuration  &  \\
0x01          & INTERRUPT ENABLE                   & 00000000                  & Interrupt Enable                       &  \\
0x02          & TEMP\_OFFSET\_ADJUST               & 00000000                  & Temperature offset adjustment          &  \\
0x03          & HUM\_OFFSET\_ADJUST                & 00000000                  & Humidity offset adjustment             &  \\
0x04          & RESET\&DRDY/INT CONF               & 00000000                  & Soft Reset and Interrupt Configuration &  \\
0x05          & MEASUREMENT CONFIGURATION  & 00000000 & Measurement Configuration                                      &  \\
0x06          & TEMPERATURE LOW                    & 00000000                  & Temperature {[}7:0{]}                  &  \\
0x07          & TEMPERATURE HIGH                   & 00000000                  & Temperature {[}15:8{]}                 &  \\
0x08          & HUMIDITY LOW                       & 00000000                  & Humidity {[}7:0{]}                     &  \\
0x09          & HUMIDITY HIGH                      & 00000000                  & Humidity {[}15:8{]}                    & 
\end{tabular}
\end{table}

\begin{table}[]
\begin{tabular}{p{0.1\linewidth}p{0.2\linewidth}p{0.1\linewidth}p{0.1\linewidth}p{0.4\linewidth}}
\multicolumn{5}{l}{Detailed Description} \\
\multicolumn{5}{l}{Address 0x00 Interrupt DRDY:} \\
\multicolumn{5}{l}{Field Descriptions} \\
BIT & FIELD        & TYPE & RESET    & DESCRIPTION                                                                                      \\
7   & DRDY\_STATUS & R/W  & 0        & DataReady bit status: 0 = Data Not Ready, 1 = Data Ready, DRDY\_STATUS is cleared to 0 when read \\
6   & RES          & 0    & Reserved &                                                                                                  \\
5   & RES          & 0    & Reserved &                                                                                                  \\
4   & RES          & 0    & Reserved &                                                                                                  \\
3   & RES          & 0    & Reserved &                                                                                                  \\
2   & RES          & 0    & Reserved &                                                                                                  \\
1   & RES          & 0    & Reserved &                                                                                                  \\
0   & RES          & 0    & Reserved &                                                                                                  \\
\\
\multicolumn{5}{p{\linewidth}}{DRDY\_STATUS indicates that temperature and/or humidity conversion is terminated. This bit is cleared when the Interrupt/DRDY register is read or the output registers TEMPERATURE\_HIGH, TEMPERATURE\_LOW, HUMIDITY\_HIGH and HUMIDITY\_LOW are read.} \\
\multicolumn{5}{p{\linewidth}}{The TL\_STATUS indicates that the Temperature Threshold LOW value is exceeded. The behavior is defined by 0x0E Configuration register value. The bit is cleared when the register Interrupt DRDY is read.} \\
\multicolumn{5}{p{\linewidth}}{The TH\_STATUS indicates that the Temperature Threshold HIGH value is exceeded. The behavior is defined by 0x0E Configuration register value. The bit is cleared when the register Interrupt DRDY is read.} \\
\multicolumn{5}{p{\linewidth}}{The HH\_STATUS indicates that the Humidity Threshold HIGH value is exceeded. The behavior is defined by 0x0E Configuration register value. The bit is cleared when the register Interrupt DRDY is read.} \\
\multicolumn{5}{p{\linewidth}}{The HL\_STATUS indicates that the Humidity Threshold LOW value is exceeded. The behavior is defined by 0x0E Configuration register value. The bit is cleared when the register Interrupt DRDY is read. } \\
\multicolumn{5}{p{\linewidth}}{DRDY/INT pin behaves like the STATUS bits based on the 0x0E Configuration register value.} \\
\end{tabular}
\end{table}

\begin{table}[]
\begin{tabular}{p{0.1\linewidth}p{0.2\linewidth}p{0.1\linewidth}p{0.1\linewidth}p{0.3\linewidth}}
\multicolumn{5}{l}{Address 0x01 Interrupt Configuration:} \\
\multicolumn{5}{l}{Field Descriptions} \\
BIT & FIELD        & TYPE & RESET    & DESCRIPTION                                                                                                     \\
7   & DRDY\_ENABLE & R/W  & 0        & DataReady Interrupt enable: 0 = DataReady Interrupt generator disable, 1 = DataReady Interrupt generator enable \\
6   & RES          & 0    & Reserved &                                                                                                                 \\
5   & RES          & 0    & Reserved &                                                                                                                 \\
4   & RES          & 0    & Reserved &                                                                                                                 \\
3   & RES          & 0    & Reserved &                                                                                                                 \\
2   & RES          & 0    & Reserved &                                                                                                                 \\
1   & RES          & 0    & Reserved &                                                                                                                 \\
0   & RES          & 0    & Reserved &                                                                                                                 \\
\end{tabular}
\end{table} 

\begin{table}[]
\begin{tabular}{p{0.1\linewidth}p{0.3\linewidth}p{0.1\linewidth}p{0.1\linewidth}p{0.3\linewidth}}
\multicolumn{5}{l}{Address 0x02 Temperature Offset Adjustment} \\
BIT       & FIELD                          & TYPE & RESET    & DESCRIPTION                                                             \\
{[}7:0{]} & TEMP\_OFFSET\_ADJUST {[}7:0{]} & R/W  & 00000000 & Temperature offset adjustment. Added to the converted Temperature value \\
\multicolumn{5}{p{\linewidth}}{The temperature can be adjusted adding the following values that are enable settings the equivalents bits:}\\
\multicolumn{5}{p{\linewidth}}{7 +40°C} \\
\multicolumn{5}{p{\linewidth}}{6 +30°C} \\
\multicolumn{5}{p{\linewidth}}{5 +20°C} \\
\multicolumn{5}{p{\linewidth}}{4 +10°C} \\
\multicolumn{5}{p{\linewidth}}{3 -10°C} \\
\multicolumn{5}{p{\linewidth}}{2 -20°C} \\
\multicolumn{5}{p{\linewidth}}{1 -30°C} \\
\multicolumn{5}{p{\linewidth}}{0 -40°C} \\
\end{tabular}
\end{table}

\begin{table}[]
\begin{tabular}{p{0.1\linewidth}p{0.3\linewidth}p{0.1\linewidth}p{0.1\linewidth}p{0.3\linewidth}}
\multicolumn{5}{l}{Address 0x03 Humidity Offset Adjustment}\\
BIT       & FIELD                        & TYPE & RESET    & DESCRIPTION                                                       \\
{[}7:0{]} & HUM\_OFFSET\_ADJUST{[}7:0{]} & R/W  & 00000000 & Humidity offset adjustment. Added to the converted Humidity value \\
\\
\multicolumn{5}{p{\linewidth}}{The humidity can be adjusted adding the following values that are enable settings the equivalents bits:} \\
\multicolumn{5}{p{\linewidth}}{7 +4\%RH} \\
\multicolumn{5}{p{\linewidth}}{6 +3\%RH} \\
\multicolumn{5}{p{\linewidth}}{5 +2\%RH} \\
\multicolumn{5}{p{\linewidth}}{4 +1\%RH} \\
\multicolumn{5}{p{\linewidth}}{3 -1\%RH} \\
\multicolumn{5}{p{\linewidth}}{2 -2\%RH} \\
\multicolumn{5}{p{\linewidth}}{1 -3\%RH} \\
\multicolumn{5}{p{\linewidth}}{0 -4\%RH} \\
\multicolumn{5}{p{\linewidth}}{The resulting humidity offset is a summation of the register bits that have been enabled (i.e. programmed to 1). Some examples:}\\
\multicolumn{5}{p{\linewidth}}{1. Programming HUM\_OFFSET\_ADJUST to 00000001 adjusts the reported humidity by -4\%RH.}\\
\multicolumn{5}{p{\linewidth}}{2. Programming HUM\_OFFSET\_ADJUST to 00000011 adjusts the reported humidity by -7\%RH.}\\
\end{tabular}
\end{table}

\noindent

\begin{table}[]
\begin{tabular}{p{0.1\linewidth}p{0.2\linewidth}p{0.1\linewidth}p{0.1\linewidth}p{0.4\linewidth}}
\multicolumn{5}{l}{Address 0x04 Reset and DRDY/INT Configuration Register:}\\
\multicolumn{5}{l}{Address 0x04 Configuration Field Descriptions}\\
BIT       & FIELD        & TYPE & RESET & DESCRIPTION                                                                                                                                              \\
7         & SOFT\_RES    & R/W  & 0     & 0 = Normal Operation mode, this bit is self-clear. 1 = Soft Reset. EEPROM value reload and registers reset                                               \\
{[}6:4{]} & AMM{[}2:0{]} & R/W  & 000   & Auto Measurement Mode (AMM). 000 = Disabled. Initiate measurement via I2C. 001 = 1/60Hz (1 samples every minute). 010:111 = 1Hz (1 samples every second) \\
3         & RES          & N/A  & 0     & Reserved                                                                                                                                                 \\
2         & DRDY/INT\_EN & R/W  & 0     & DRDY/INT\_EN pin configuration: 0 = High Z, 1 = Enable                                                                                                   \\
1         & INT\_POL     & R/W  & 0     & Interrupt polarity: 0 = Active Low, 1 = Active High                                                                                                      \\
0         & INT\_MODE    & R/W  & 0     & Interrupt mode: 0 = Level sensitive, 1 = Comparator mode                                                                                          
\end{tabular}
\end{table}

\begin{table}[]
\begin{tabular}{p{0.1\linewidth}p{0.2\linewidth}p{0.1\linewidth}p{0.1\linewidth}p{0.4\linewidth}}
\multicolumn{5}{l}{Address 0x05 Measurement Configuration:}\\
\multicolumn{5}{l}{Address 0x05 Measurement Configuration Field Descriptions}\\
BIT & FIELD               & TYPE & RESET & DESCRIPTION                                                                                            \\
7:6 & TRES{[}1:0{]}       & R/W  & 00    & Temperature resolution - 00: 14 bit, 01: 11 bit, 10: 9 bit, 11: NA                                     \\
5:4 & HRES{[}1:0{]}       & R/W  & 00    & Humidity resolution - 00: 14 bit, 01: 11 bit, 10: 9 bit, 11: NA                                        \\
3   & RES                 & R/W  & 0     & Reserved                                                                                               \\
2:1 & MEAS\_CONF{[}1:0{]} & R/W  & 00    & Measurement configuration - 00: Humidity + Temperature, 01: Temperature only, 10: NA, 11: NA           \\
0   & MEAS\_TRIG          & R/W  & 0     & Measurement trigger - 0: no action, 1: Start measurement. Self-clearing bit when measurement completed \\
0   & INT\_MODE           & R/W  & 0     & Interrupt mode: 0 = Level sensitive, 1 = Comparator mode \\
\\
\multicolumn{5}{p{\linewidth}}{Address 0x06 Temperature LSB \& Address 0x07 Temperature MSB: The temperature register is a 16-bit result register in binary format (the 2 LSBs D1 and D0 are always 0). The result of the acquisition is always a 14-bit value, while the resolution is related to one selected in Measurement Configuration register. The temperature can be calculated from the output data with Equation:
Temperature (Celsius) = (Temperature [15 : 0] / 2\^16) * 100 - 314} \\
\\
\multicolumn{5}{p{\linewidth}}{Address 0x08 Humidity LSB \& Address 0x09 Humidity MSB: The humidity register is a 16-bit result register in binary format (the 2 LSBs D1 and D0 are always 0). The result of the acquisition is always a 14-bit value, while the resolution is related to one selected in Measurement Configuration register. The humidity can be calculated from the output data with the Equation:
Humidity (\%RH) = (Humidity [15 : 0] / 2\^16) * 14}
\end{tabular}
\end{table}
\end{document}